\documentclass[11pt,twoside]{article}
\usepackage{atmp}
\copyrightnotice{2003}{7}{233}{268}	

\def\be{\begin{equation}}
\def\ee{\end{equation}}
\def\ba{\begin{eqnarray}}
\def\ea{\end{eqnarray}}
\def\g{\gamma}

\def\A{{\cal A}}

\def\H{{\cal H}}

\def\S{{\cal S}}

\def\Ab{{\bar \A}}

\def\Cyl{{\rm Cyl}}
\def\cyl{\Cyl}

\def\SU{{\rm SU}}

\def\a{\alpha}
\def\g{\gamma}
\def\e{\epsilon}

\def\lp{{\ell}_{\rm Pl}}

\def\j{{\bf j}}

\def\M{M}

\def\q{{}^o\!q}
\def\e{{}^o\!e}
\def\w{{}^o\!\omega}

\def\L{\mu}
\def\ar{A_{S,f}}

\def\f{\frac}

\begin{document}

\title{Mathematical structure of\\ loop quantum cosmology}

\url{gr-qc/0304074}		

\author{Abhay Ashtekar${}^{1,2}$, Martin Bojowald${}^{1,2}$, and\\ Jerzy Lewandowski${}^{3,1,2}$}		
\address{1. Center for Gravitational Physics and Geometry,\\
Physics Department, Penn State, University Park, PA 16802, USA\\
2. Erwin Schr\"odinger Institute, Boltzmanngasse 9, 1090 Vienna,
Austria\\
3. Institute of Theoretical Physics, University of Warsaw, ul.
Ho\.{z}a 69, 00-681 Warsaw, Poland}

\markboth{\it Mathematical structure of loop quantum cosmology}{\it A.\ Ashtekar, M.\ Bojowald and J.\ Lewandowski}

\begin{abstract}
Applications of Riemannian quantum geometry to cosmology have had
notable successes. In particular, the fundamental discreteness
underlying quantum geometry has led to a natural resolution of the
big bang singularity. However, the precise mathematical structure
underlying loop quantum cosmology  and the sense in which it
implements the full quantization program in a symmetry reduced
model has not been made explicit. The purpose of this paper is to
address these issues, thereby providing a firmer mathematical and
conceptual foundation to the subject.
\end{abstract}
%
%
%

\section{Introduction}
\label{s1}

In cosmology, one generally freezes all but a finite number of
degrees of freedom by imposing spatial homogeneity (and sometimes
also isotropy). Because of the resulting mathematical
~simplifications, the framework pro- \newline
vides a simple ~arena to ~test
ideas and ~constructions ~introduced in~the~full 

\vfill
\cutpage

\noindent
theory both at the
classical and the quantum levels. Moreover, in the classical
regime, the symmetry reduction captures the large scale dynamics
of the universe as a whole quite well. Therefore, in the quantum
theory, it provides a useful test-bed for analyzing the important
issues related to the fate of classical singularities.

Over the last three years, ramifications of Riemannian quantum
geometry to cosmology have been investigated systematically.  First,
already at the kinematic level it was found that, thanks to the
fundamental discreteness of quantum geometry, the inverse scale factor
---and hence also the curvature--- remains bounded on the kinematical
Hilbert space \cite{mb1}. Second, while classical dynamics is
described by differential equations, the quantum Hamiltonian
constraint can be interpreted as providing a difference equation for
the `evolution' of the quantum state \cite{mb2}. Furthermore, all
quantum states remain regular at the classical big-bang; one can
`evolve' right through the point at which classical physics stops
\cite{mb3}. Third, the Hamiltonian constraint together with the
requirement ---called pre-classicality--- that the universe be
classical at late times severely restricts the quantum state and, in
the simplest models, selects the state uniquely \cite{mb4}. There are
also phenomenological models which allow us to study simple effects of
quantum geometry leading to a behavior qualitatively different from
the classical one \cite{mb4a}. Finally, the qualitative features are
robust \cite{mb5} and extend also to more complicated cosmological
models \cite{mb6}. These results are quite surprising from the
perspective of the `standard' quantum cosmology which was developed in
the framework of geometrodynamics and, together, they show that, once
the quantum nature of geometry is appropriately incorporated, the
physical predictions change qualitatively in the Planck era.

In spite of these striking advances, the subject has remained
incomplete in several respects. First, in the existing treatments,
certain subtleties which turn out to have important ramifications
were overlooked and the underlying mathematical structure was
somewhat oversimplified. This sometimes led to the impression that
some of the physically desirable but surprising results arose
simply because of ad-hoc assumptions. Second, the essential
reasons why loop quantum cosmology is so different from the
`standard' quantum cosmology have not been spelled out. Third,
while it is clear that the key constructions and techniques used
in loop quantum cosmology are inspired by those developed in the
full theory based on quantum geometry
\cite{ai,al2,jb1,mm,al3,al4,tt,al8}, the parallels and the
differences between the full theory and the symmetry reduced
models have not been discussed in detail. In this paper, we will
address these issues, providing a sounder foundation for the
striking results obtained so far. The paper also has a secondary,
pedagogical goal: it will also provide an introduction to quantum
geometry and loop quantum gravity for readers who are not familiar
with these areas. Our discussion should significantly clarify the
precise mathematical structure underlying loop quantum cosmology
and its relation to the full theory as well as to
geometrodynamical quantum cosmology. However, we will not address
the most important and the most difficult of open issues: a
systematic \emph{derivation} of loop quantum cosmology from full
quantum gravity.

Results of quantum cosmology often provide important qualitative
lessons for full quantum gravity. However, while looking for these
lessons, it is important to remember that the symmetry reduced
theory used here differs from the full theory in conceptually
important ways. The most obvious difference is the reduction from
a field theory to a mechanical system, which eliminates the
potential ultra-violet and infra-red problems of the full theory.
In this respect the reduced theory is much simpler. However, there
are also two other differences ---generally overlooked--- which
make it conceptually and technically \emph{more} complicated, at
least when one tries to directly apply the techniques developed
for the full theory. First, the reduced theory is usually treated
by gauge fixing and therefore fails to be diffeomorphism
invariant. As a result, key simplifications that occur in the
treatment of full quantum dynamics \cite{tt} do not carry over.
Therefore, in a certain well-defined sense, the non-perturbative
dynamics acquires \emph{new} ambiguities in the reduced theory!
The second complication arises from the fact that spatial
homogeneity introduces distant correlations. Consequently, at the
kinematical level, quantum states defined by holonomies along with
distinct edges and triad operators smeared on distinct 2-surfaces
are no longer independent. We will see that both these features
give rise to certain complications which are \emph{not} shared by
the full theory.

The remainder of this paper is divided in to four sections. In the
second, we discuss the phase space of isotropic, homogeneous
cosmologies; in the third, we construct the quantum kinematic
framework; in the fourth we impose the Hamiltonian constraint and
discuss properties of its solutions and in the fifth we summarize
the results and discuss some of their ramifications.

\section{Phase space}
 \label{s2}

For simplicity, we will restrict ourselves to spatially flat,
homogeneous, isotro\-pic cosmologies, so that the spatial isometry
group $\S$ will be the Euclidean group. Then the 3-dimensional
group ${T}$ of translations (ensuring homogeneity) acts simply and
transitively on the 3-manifold $\M$. Therefore, $\M$ is
topologically $R^3$. Through the Cartan-Killing form on the
Lie-algebra of the rotation group, the Lie algebra of translations
acquires an equivalence class of positive definite metrics related
by an overall constant. Let us fix a metric in this class and an
action of the Euclidean group on $M$. This will endow $\M$ with a
fiducial flat metric $\q_{ab}$. Finally, let us fix a constant
orthonormal triad $\e^a_i$ and a co-triad $\w_a^i$ on $\M$,
compatible with $\q_{ab}$.

Let us now turn to the gravitational phase space in the connection
variables. In the full theory, the phase space consists of pairs
$(A_a^i,\, E^a_i)$ of fields on a 3-manifold $\M$, where $A_a^i$
is an $\SU(2)$ connection and $E^a_i$ a triplet of vector fields
with density weight 1 \cite{aa1}. (The density weighted
orthonormal triad is given by $\g E^a_i$, where $\g$ is the
Barbero-Immirzi parameter.) Now, a pair $(A^\prime_a{}^i,\,
E^{\prime a}_i)$ on $\M$ will be said to be \emph{symmetric} if
for every $s\in \S$ there exists a local gauge transformation $g:
\M\rightarrow SU(2)$, such that
\be \label{ge} (s^*A^\prime, \, s^*E^\prime)\ =\ (g^{-1}A'g\, + \,
g^{-1}dg,\,\, g^{-1}E'g). \end{equation}
As is usual in cosmology, we will fix the local diffeomorphism and
gauge freedom. To do so, note first that for every symmetric pair
$(A^\prime,\, E^\prime)$ (satisfying the Gauss and diffeomorphism
constraints) there exists an unique equivalent pair $(A,\, E)$
such that
\be\label{ss} A\ =\ \tilde{c}\,\, \w^{i} \tau_i,\quad {E}\ =\
\tilde{p}\, \sqrt{\q}\,\,\, \e_{i}\tau^i \end{equation}
where $\tilde{c}$ and $\tilde{p}$ are constants, carrying the only
non-trivial information contained in the pair $(A^\prime,
E^\prime)$, and the density weight of ${E}$ has been absorbed in
the determinant of the fiducial metric. (Our conventions are such
that $[\tau_i, \tau_j] = \epsilon_{ijk}\tau^k$, i.e., $2i \tau_i =
\sigma_i$, where $\sigma_i$ are the Pauli matrices.)

In terms of $\tilde{p}$, the physical orthonormal triad $e^a_i$
and its inverse $e_a^i$ (both of zero density weight) are given
by:
\be  e^a_i \equiv \g \tilde{p}\, \sqrt{\frac{\q}{q}}\,\,\, \e^a_i
= ({\rm sgn} \tilde{p})\, |\g \tilde{p}|^{-\frac{1}{2}} \,\,
\e^a_i, \quad {\rm and} \quad e_a^i = ({\rm sgn} \tilde{p})\,\,\,
|\g \tilde{p}|^{\frac{1}{2}}\,\, \w_a^i \end{equation}
where $q=\det q_{ab}= |\det\gamma E^a_i|$,\,\, ${\rm sgn}$ stands
for the sign function and $\g$ is the Barbero-Immirzi parameter.
As in the full theory, the Barbero-Immirzi parameter $\g$ and the
determinant factors are necessary to convert the (density
weighted) momenta $E^a_i$ in to geometrical (unweighted) triads
$e^a_i$ and co-triads $e_a^i$. The sign function arises because
the connection dynamics phase space contains triads with both
orientations and, because we have fixed a fiducial triad $\e^a_i$,
the orientation of the physical triad $e^a_i$ changes with the
sign of $\tilde{p}$. (As in the full theory, we also allow
degenerate co-triads which now correspond to $\tilde{p}=0$, for
which the triad vanishes.)

Denote by $\A_S$ and $\Gamma^S_{\rm grav}$ the subspace of the
gravitational configuration space $\A$ and of the gravitational
phase space ${\Gamma_{\rm grav}}$ defined by (\ref{ss}). Tangent
vectors $\delta$ to $\Gamma^S_{\rm grav}$ are of the form:
\be \label{tv} \delta = (\delta A,\, \delta E), \quad {\rm with}
\quad \delta A_a^i \equiv (\delta \tilde{c})\,\, \w_a^i, \,\,\,
\delta E_i^a \equiv (\delta \tilde{p})\,\, \e^a_i . \end{equation}
Thus, $\A_S$ is 1-dimensional and  $\Gamma^S_{\rm grav}$ is
2-dimensional: we made a restriction to \emph{symmetric} fields
and solved and gauge-fixed the gauge and the diffeomorphism
constraints, thereby reducing the local, infinite number of
gravitational degrees of freedom to just one.

Because $\M$ is non-compact and our fields are spatially
homogeneous, various integrals featuring in the Hamiltonian
framework of the full theory diverge. This is in particular the
case for the symplectic structure of the full theory:
\be {\Omega}_{\rm grav}(\delta_1, \, \delta_2)\,  = \,
\frac{1}{8\pi\gamma G} \int_{\M} d^3x \left(\delta_1
A^i_a(x)\delta_2 E_i^a(x) - \delta_2 A^i_a(x)\delta_1
E_i^a(x)\right)\,  .\end{equation}
However, the presence of spatial homogeneity enables us to bypass
this problem in a natural fashion: Fix a `cell' ${\cal V}$ adapted
to the fiducial triad and restrict all integrations to this cell.
(For simplicity, we will assume that this cell is cubical with
respect to $\q_{ab}$.) Then the gravitational symplectic structure
$\Omega_{\rm grav}$ on $\Gamma_{\rm grav}$ is given by:
\be {\Omega}_{\rm grav}(\delta_1, \, \delta_2)\,  = \,
\frac{1}{8\pi\gamma G} \int_{\cal V} d^3x \left(\delta_1
A^i_a(x)\delta_2 E_i^a(x) - \delta_2 A^i_a(x)\delta_1
E_i^a(x)\right)\,  .\end{equation}
Using the form (\ref{tv}) of the tangent vectors, the pull-back of
${\Omega}$ to $\Gamma^S_{\rm grav}$ reduces just to:
\be \Omega^S_{\rm grav}\, =\, \frac{3V_o}{8\pi \gamma G} \,\,
d\tilde{c} \wedge d\tilde{p} \end{equation}
where $V_o$ is the volume of ${\cal V}$ with respect to the
auxiliary metric $\q_{ab}$. (Had $M$ been compact, we could set
${\cal V} = M$ and $V_o$ would then be the total volume of $M$
with respect to $\q_{ab}$.) Thus, we have specified the
gravitational part of the reduced phase space. We will not need to
specify matter fields explicitly but only note that, upon similar
restriction to symmetric fields and fixing of gauge and
diffeomorphism freedom, we are led to a finite dimensional phase
space also for matter fields.

In the passage from the full to the reduced theory, we introduced a
fiducial metric $\q_{ab}$. There is a freedom in rescaling this metric
by a constant: $\q_{ab} \mapsto k^2 \q_{ab}$. Under this rescaling the
canonical variables $\tilde{c}, \tilde{p}$ transform via $\tilde{c}
\mapsto k^{-1} \tilde{c}$ and $\tilde{p} \mapsto k^{-2}
\tilde{p}$. (This is analogous to the fact that the scale factor
$\tilde{a}=\sqrt{|\tilde{p}|}$ in geometrodynamics rescales by a
constant under the change of the fiducial flat metric.) Since
rescalings of the fiducial metric do not change physics, by themselves
$\tilde{c}$ and $\tilde{p}$ do not have direct physical
meaning. Therefore, it is convenient to introduce new variables:
\be  c = V_o^{\f{1}{3}} \tilde{c} \quad {\rm and} \quad
     p = V_o^{\f{2}{3}} \tilde{p} \end{equation}
which are independent of the choice of the fiducial metric
$\q_{ab}$. In terms of these, the symplectic structure is given by
\be \label{qcsym} \Omega^S_{\rm grav}\, =\, \frac{3}{8\pi \gamma
G} \,\, dc \wedge dp\, ; \end{equation}
it is now independent of the volume $V_o$ of the cell ${\cal V}$
and makes no reference to the fiducial metric. In the rest of the
paper, we will work with this phase space description. Note that
now the configuration variable $c$ is dimensionless while the
momentum variable $p$ has dimensions $({\rm length})^2$. (While
comparing results in the full theory, it is important to bear in
mind that these dimensions are different from those of the
gravitational connection and the triad there.) In terms of $p$,
the physical triad and co-triad are given by:
\be \label{e1} e^a_i = ({\rm sgn}\,p) |\g p|^{-\frac{1}{2}}\,\,\,
(V_o^{\frac{1}{3}}\,  \e^a_i), \quad {\rm and} \quad e_a^i = ({\rm
sgn}\, {p}) |\g p|^{\frac{1}{2}}\,\,\, (V_o^{- \frac{1}{3}}\,
\w_a^i) \end{equation}

Finally, let us turn to constraints. Since the Gauss and the
diffeomorphism constraints are already satisfied, there is a
single non-trivial Scalar/Ha\-miltonian constraint (corresponding to
a constant lapse):
\be \label{scalar} -\frac{6}{\gamma^2}\,\, c^2\, {\rm sgn}p\,
\sqrt{|p|} \, + \, 8\pi G\, C_{\rm matter}\ =\ 0\, . \end{equation}
%

\section{Quantization: Kinematics}
\label{s3}

\subsection{Elementary variables}
\label{s3.1}

Let us begin by singling out `elementary functions' on the
classical phase space which are to have unambiguous quantum
analogs. In the full theory, the configuration variables are
constructed from holonomies $h_e(A)$ associated with edges $e$ and
momentum variables, from $E(S,f)$, momenta $E$ smeared with test
fields $f$ on 2-surfaces \cite{al4,al5,almmt,al8}. But now,
because of homogeneity and isotropy, we do not need all edges $e$
and surfaces $S$. Symmetric connections $A$ in $\A_S$ can be
recovered knowing holonomies $h_e$ along edges $e$ which lie along
straight lines in $M$. Similarly, it is now appropriate to smear
triads only across squares (with respect to
$\q_{ab}$).%
\footnote{Indeed, we could just consider edges lying in a single
straight line and a single square. We chose not to break the
symmetry artificially and consider instead all lines and all
squares.}

The $\SU(2)$ holonomy along an edge $e$ is given by:
\be \label{hol} h_e(A) := {\cal P}\, \exp {\int_e\, A} = \cos
\f{\L c}{2} + 2\, \sin \f{\L c}{2}\,\, (\dot{e}^a\w_a^i)\, \tau^i
\end{equation}
where $\L \in (-\infty,\, \infty)$ (and $\L V_o^{\f{1}{3}}$ is
the oriented length of the edge with respect to $\q_{ab}$).
Therefore, the algebra generated by sums of products of matrix
elements of these holonomies is just the algebra of \emph{almost
periodic functions} of $c$, a typical element of which can be
written as:
\be \label{f} g(c) = \sum_j \xi_j \,\, e^{i\f{\L_j c}{2}} \, \end{equation}
where $j$ runs over a finite number of integers (labelling edges),
$\L_j \in R$ and $\xi_j \in C$. In the terminology used in the
full theory, one can regard a finite number of edges as providing
us with a graph (since, because of homogeneity, the edges need not
actually meet in vertices now) and the function $g(A)$ as a
cylindrical function with respect to that graph. The vector space
of these almost periodic functions is, then, the analog of the
space  $\Cyl$ of cylindrical functions on $\A$ in the full theory
\cite{al2,jb1,mm,al4,almmt}. We will call it the space of
cylindrical functions of \emph{symmetric} connections and denote
it by $\Cyl_S$.

In the full theory, the momentum functions $E(S,f)$ are obtained
by smearing the `electric fields' $E^a_i$ with an ${\rm
su}(2)$-valued function $f^i$ on a 2-surface $S$. In the
homogeneous case, it is natural to use constant test functions
$f^i$ and let $S$ be squares tangential to the fiducial triad
$\e^a_i$. Then, we have:
\be E(S, f) = \int_S \Sigma^i_{ab}f_i dx^a dx^b = p\,\,
V_o^{-\f{2}{3}} \ar \end{equation}
where $\Sigma^i_{ab} = \eta_{abc} E^{ci}$ and where $\ar$ equals
the area of $S$ as measured by $\q_{ab}$, times an obvious
orientation factor (which depends on $f_i$).
Thus, apart from a kinematic factor determined by the background
metric, the momenta are given just by $p$. In terms of classical
geometry, $p$ is related to the \emph{physical} volume of the
elementary cell ${\cal V}$ via
\be V = |p|^{\f{3}{2}} \, .\end{equation}
Finally, the only non-vanishing Poisson bracket between these
elementary functions is:
\be \{g(A), p\} = \frac{8\pi\g G}{6} \sum_j (i\L_j \xi_j)\,
e^{i\f{\L_j c}{2}}\, .\end{equation}
%
%
%
Since the right side is again in $\Cyl_S$, the space of elementary
variables is closed under the Poisson bracket. Note that, in
contrast with the full theory, now the smeared momenta $E(S,f)$
commute with one another since they are all proportional to $p$
because of homogeneity and isotropy. \emph{This implies that now
the triad representation does exist.} In fact it will be
convenient to use it later on in this paper.

\subsection{Representation of the algebra of elementary variables}
\label{s3.2}

To construct quantum kinematics, we seek a representation of this
algebra of elementary variables. In the full theory, one can use
the Gel'fand theory to first find a representation of the
$C^\star$ algebra $\Cyl$ of configuration variables and then
represent the momentum operators on the resulting Hilbert space
\cite{ai,al2,al4,almmt}. In the symmetry reduced model, we can
follow the same procedure. We will briefly discuss the abstract
construction and then present the explicit Hilbert space and
operators in a way that does not require prior knowledge of the
general framework.

Let us begin with the $C^\star$ algebra $\Cyl_S$ of almost
periodic functions on $\A_S$ which is topologically $R$. The
Gel'fand theory now guarantees that there is a compact Hausdorff
space $\bar{R}_{\rm Bohr}$, the algebra of \emph{all} continuous
functions on which is isomorphic with $\Cyl_S$. $\bar{R}_{\rm
Bohr}$ is called the \emph{Bohr compactification of the real line}
$\A_S$, and $\A_S$ is densely embedded in it. The Gel'fand theory
also implies that the Hilbert space is necessarily
$L^2(\bar{R}_{\rm Bohr}, d\mu)$ with respect to a regular Borel
measure $\mu$. Thus, the classical configuration space $\A_S$ is
now extended to the quantum configuration space $\bar{R}_{\rm
Bohr}$. The extension is entirely analogous to the extension from
the space $\A$ of smooth connections to the space $\Ab$ of
generalized connections in the full theory \cite{ai,al2,al4,al8}
and came about because, as in the full theory, our configuration
variables are constructed from holonomies. In the terminology used
in the full theory, elements $\bar{c}$ of $\bar{R}_{\rm Bohr}$ are
`generalized symmetric connections'. In the full theory, $\Ab$ is
equipped with a natural, faithful, `induced' Haar measure, which
enables one to construct the kinematic Hilbert space and a
preferred representation of the algebra of holonomies and smeared
momenta
\cite{al2,jb1,mm,al3,al4}.%
\footnote{Recently, this representation has been shown to be
uniquely singled out by the requirement of diffeomorphism
invariance \cite{hs,lo,st}.}
Similarly, $\bar{R}_{\rm Bohr}$ is equipped with a natural
faithful, `Haar measure' which we will denote
by $\mu_o$.%
\footnote{$\bar{R}_{\rm Bohr}$ is a compact Abelian group and
$d\mu_o$ is the Haar measure on it. In non-relativistic quantum
mechanics, using $\bar{R}_{\rm Bohr}$ one can introduce a new
representation of the standard Weyl algebra. It is
\emph{inequivalent} to the standard Schr\"odinger representation
and naturally incorporates the idea that spatial geometry is
discrete at a fundamental scale. Nonetheless, it reproduces the
predictions of standard Schr\"odinger quantum mechanics within its
domain of validity. (For details, see \cite{afw}). There is a
close parallel with the situation in quantum cosmology, where the
role of the Schr\"odinger representation is played by `standard'
quantum cosmology of geometrodynamics.}

Let us now display all this structure more explicitly. The Hilbert
space $\H_{\rm grav}^S = L^2(\bar{R}_{\rm Bohr}, d\mu_o)$ can be
made `concrete' as follows. It is the Cauchy completion of the
space $\Cyl_S$ of almost periodic functions of $c$ with respect to
the inner product:
\be \label{ip} \langle{e^{i\f{\L_1 c}{2}}}|{e^{i\f{\L_2 c}{2}}}
\rangle = \delta_{\L_1,\L_2} \end{equation}
(Note that the right side is the Kronecker delta, not the Dirac
distribution.) Thus, the almost periodic functions ${\cal N}_\L
(c) := e^{i\L c/2}$ constitute an orthonormal basis in $\H_{\rm
grav}^S$. $\Cyl_S$ is dense in $\H_{\rm grav}^S$, and serves as a
common domain for all elementary operators. The configuration
variables act in the obvious fashion: For all $g_1$ and $g_2$ in
$\Cyl_S$, we have:
\be (\hat{g}_1 g_2 )({c}) = g_1({c}) g_2({c}) \end{equation}
Finally, we represent the momentum operator via
\be \hat{p} = -i \f{\g\lp^2}{3}\, \frac{d}{dc}, \quad {\rm
whence,}\quad (\hat{p} g)(c) = \f{\g\lp^2}{6}\,\, \sum_j [\xi_j
\L_j]\,\, {\cal N}_{\L_j} \, \end{equation}
where $g \in \Cyl_S$ is given by (\ref{f}) and, following
conventions of loop quantum cosmology, we have set $\lp^2 = 8\pi
G\hbar$. (Unfortunately, this convention is different from that
used in much of quantum geometry where $G\hbar$ is set equal to
$\lp$.)

As in the full theory, the configuration operators are bounded,
whence their action can be extended to the full Hilbert space
$\H^S_{\rm grav}$, while the momentum operators are unbounded but
essentially self-adjoint. The basis vectors ${\cal N}_\L$ are
normalized eigenstates of $\hat{p}$. As in quantum mechanics, let
us use the bra-ket notation and write ${\cal N}_\L (c) = \langle
c|\L\rangle$. Then,
\be \hat{p}\, |\L\rangle \, = \, \f{\L\gamma\lp^2}{6}\,\, |\L
\rangle \, \equiv \, p_\mu\, |\L\rangle \, . \end{equation}
Using the relation $V = |p|^{3/2}$ between $p$ and physical volume
of the cell ${\cal V}$ we have:
\be \label{vol} \hat{V}\, |\L\rangle =  \left( \frac{\gamma
|\L|}{6}\right)^{\f{3}{2}}\,\, \lp^3\, |\L\rangle\, \equiv V_\L\,
|\L\rangle . \end{equation}
This provides us with a physical meaning of $\L$: apart from a
fixed constant, $|\L|^{3/2}$ is the \emph{physical} volume of the
cell ${\cal V}$ \emph{in Planck units}, when the universe is in
the quantum state $|\L\rangle$. Thus, in particular, while the
volume $V_o $ of the cell ${\cal V}$ with respect to the fiducial
metric $\q_{ab}$ may be `large', its physical volume in the
quantum state $|\L =1\rangle$ is $(\g/6)^{3/2} \lp^3$. This fact
will be important in sections \ref{s3.3} and \ref{s4.1}.

Note that the construction of the Hilbert space and the
representation of the algebra is entirely parallel to that in the
full theory. In particular, $\Cyl_S$ is analogous to $\Cyl$;
$\bar{R}_{\rm Bohr}$ is analogous to $\Ab$;\, ${\cal N}_\L$ to the
spin network states ${\cal N}_{\a, \j,{\bf I}}$ (labelled by a
graph $g$ whose edges are assigned half integers $\j$ and whose
vertices are assigned intertwiners ${\bf I}$\cite{rs,jb2,almmt}).
$\hat{g}$ are the analogs of configuration operators defined by
elements of $\Cyl$ and  $\hat{p}$ is analogous to the triad
operators. %
In the full theory, holonomy operators are well-defined but there
is no operator representing the connection itself. Similarly,
$\hat{\cal N}_\L$ are well defined unitary operators on $\H^S_{\rm
grav}$ but they fail to be continuous with respect to $\L$, whence
there is no operator corresponding to $c$ on $\H_{\rm grav}^S$.
Thus, to obtain operators corresponding to functions on the
gravitational phase space $\Gamma^S_{\rm grav}$ we have to
\emph{first express them in terms of our elementary variables
${\cal N}_\L$ and $p$ and then promote those expressions to the
quantum theory.} Again, this is precisely the analog of the
procedure followed in the full theory.

There is, however, one important difference between the full and
the reduced theories: while eigenvalues of the momentum (and other
geometric) operators in the full theory span only a discrete
subset of the real line, now every real number is a permissible
eigenvalue of $\hat{p}$. This difference can be directly
attributed to the high degree of symmetry. In the full theory,
eigenvectors are labelled by a pair $(e, j)$ consisting of
continuous label $e$ (denoting an edge) and a discrete label $j$
(denoting the `spin' on that edge), and the eigenvalue is dictated
by $j$. Because of homogeneity and isotropy, the pair $(e,j)$ has
now collapsed to a single continuous label $\L$. Note however that
there \emph{is} a weaker sense in which the spectrum is discrete:
all eigenvectors are \emph{normalizable}. Hence the Hilbert space
can be expanded out as a direct \emph{sum} ---rather than a direct
integral--- of the 1-dimensional eigenspaces of $\hat{p}$; i.e.,
the decomposition of identity on $\H_S$ is given by a (continuous)
sum
\be I = \sum_{\L}\, |\L\rangle \langle \L| \end{equation}
rather than an integral. Although weaker, this discreteness is
nonetheless important both technically and conceptually. In the
next sub-section, we present a \emph{key} illustration.

We will conclude with two remarks.

i) In the above discussion we worked with $c,p$ rather than the
original variables $\tilde{c}, \tilde{p}$ to bring out the
physical meaning of various objects more directly. Had we used the
tilde variables, our symplectic structure would have involved
$V_o$ and we would have had to fix $V_o$ prior to quantization.
The Hilbert space and the representation of the configuration
operators would have been the same for all choices of $V_o$.
However, the representation of the momentum operators would have
changed from one $V_o$ sector to another: A change $V_o \mapsto
k^3 V_o$ would have implied $\hat{p} \mapsto k^{-2} \hat{p}$. The
analogous transformation is not unitarily implementable in
Schr\"odinger quantum mechanics nor in full quantum gravity.
However, somewhat surprisingly, it \emph{is} unitarily
implementable in the reduced model.%
\footnote{The difference from Schr\"odinger quantum mechanics can
be traced back to the fact that the eigenvectors $|p\rangle$ of
the Schr\"odinger momentum operator satisfy $\langle p| p'\rangle
= \delta(p,p')$ while the eigenvectors of our $\hat{p}$ in the
reduced model satisfy $\langle \L |\L'\rangle = \delta_{\L, \L'}$,
the Dirac delta distribution being replaced by the Kronecker
delta. In full quantum gravity, one also has the Kronecker-delta
normalization for the eigenvectors of triad (and other
geometrical) operators. Now the difference arises because there
the eigenvalues form a discrete subset of the real line while in
the symmetry reduced model they span the full line.}
Therefore, quantum physics does not change with the change of
$V_o$.  Via untilded variables, we chose to work with an unitarily
equivalent representation which does not refer to $V_o$ at all.

ii) For simplicity of presentation, in the above discussion we
avoided details of the Bohr compactification and worked with its
dense space $\Cyl_S$ instead. In terms of the compactification,
the situation can be summarized as follows. After Cauchy
completion, each element of $\H^S_{\rm grav}$ is represented by a
square-integrable function $f(\bar{c})$ of generalized symmetric
connections. By Gel'fand transform, every element $g$ of $\Cyl_S$
is represented by a function $\check{g}(\bar{c})$ on $\Cyl_S$ and
the configuration operators act via multiplication on the full
Hilbert space: $(\hat{g}_1g_2)(\bar{c}) = \check{g}_1(\bar{c})
g_2(\bar{c})$. The momentum operator $\hat{p}$ is essentially
self-adjoint on the domain consisting of the image of $\Cyl_S$
under the Gel'fand transform.

\subsection{Triad operator}
\label{s3.3}

In the reduced classical theory, curvature is simply a multiple of
the \emph{inverse} of the square of the scale factor $a=\sqrt{|p|}$.
Similarly, the matter Hamiltonian invariably involves a term
corresponding to an \emph{inverse} power of $a$. Therefore, we
need to obtain an operator corresponding to the inverse scale
factor, or the triad (with density weight zero) of (\ref{e1}). In
the classical theory, the triad coefficient diverges at the big
bang and a key question is whether quantum effects `tame' the big
bang sufficiently to make the triad operator (and hence the
curvature and the matter Hamiltonian) well behaved there.

Now, in non-relativistic quantum mechanics, the spectrum of the
operator $\hat{r}$ is the positive half of the real line, equipped
with the standard Lesbegue measure, whence the operator
$1/\hat{r}$ is a densely-defined, self-adjoint operator. By
contrast, since $\hat{p}$ admits a \emph{normalized} eigenvector
$|\L=0\rangle$ with zero eigenvalue, the naive expression of the
triad operator fails to be densely defined on $\H^S_{\rm grav}$.
One could circumvent this problem in the reduced model in an
ad-hoc manner by just making up a definition for the action of the
triad operator on $|\L=0\rangle$. But then the result would have
to be considered as an artifact of a procedure expressly invented
for the model and one would not have any confidence in its
implications for the big bang. Now, as one might expect, a similar
problem arises also in the full theory. There, a mathematically
successful strategy to define the required operators already exits
\cite{tt}: One first re-expresses the desired, potentially
`problematic' phase space function as a regular function of
elementary variables and the volume function and then replaces
these by their well-defined quantum analogs. It is appropriate to
use the same procedure also in quantum cosmology; not only is this
a natural approach but it would also test the general strategy. As
in the general theory, therefore, we will proceed in two steps. In
the first, we note that, on the reduced phase space $\Gamma_{\rm
grav}^S$, the triad coefficient ${\rm sgn}\, p\,
|p|^{-\frac{1}{2}}$ can be expressed as the Poisson bracket
$\{c,\, V^{{1}/{3}} \}$ which can be replaced by $i\hbar$ times
the commutator in quantum theory. However, a second step is
necessary because there is no operator $\hat{c}$ on $\H^S_{\rm
grav}$ corresponding to $c$: one has to re-express the Poisson
bracket in terms of \emph{holonomies} which do have unambiguous
quantum analogs.

Indeed, on $\Gamma_{\rm grav}^S$, we have:
\be \label{e2} \frac{{\rm sgn}(p)}{\sqrt{|p|}}\, = \,
\frac{4}{8\pi G \g } {\rm tr}\, \left(\sum_i\tau^i h_i\{h_i^{-1},
V^{\frac{1}{3}}\} \right)\, , \end{equation}
where
\be h_i:= {\cal P}\, \exp\left(\smallint_0^{V_0^{\frac{1}{3}}}
\e^a_iA_a^j\tau_j dt\right)\, =\,  \exp( c\tau_i)\, =\,
\cos\frac{c}{2}+ 2\sin\frac{c}{2}\tau_i \end{equation}
is the holonomy (of the connection $A_a^i$) evaluated along an
edge along the elementary cell ${\cal V}$ (i.e., an edge parallel
to the triad vector $\e^a_i$ of length $ V_o^{1/3}$ with respect
to the \emph{fiducial} metric $\q_{ab}$), and where we have summed
over $i$ to avoid singling out a specific triad vector.%
\footnote{Note that, because of the factors $h_i$ and $h_i^{-1}$
in this expression, the length of the edge is actually irrelevant.
For further discussion, see the remark at the end of this
sub-section.}
We can now pass to quantum theory by replacing the Poisson
brackets by commutators. This yields the triad (coefficient)
operator:
\ba \label{qtriad1} \widehat{\left[{\frac{{\rm
sgn}(p)}{\sqrt{|p|}}}\right]}\, &=&\, -\frac{4i}{\gamma\lp^2} {\rm
tr}\left(\sum_i\tau^i \hat{h}_i [\hat{h}_i^{-1},
  \hat{V}^{\frac{1}{3}}]\right)\nonumber\\
&=&\, -\frac{12i}{\gamma\lp^2} \left(\sin\frac{c}{2}
\hat{V}^{\frac{1}{3}} \cos\frac{c}{2}- \cos\frac{c}{2}
\hat{V}^{\frac{1}{3}} \sin\frac{c}{2}\right) \ea

Although this operator involves both configuration and momentum
operators, it commutes with $\hat{p}$, whence its eigenvectors are
again $|\L\rangle$. The eigenvalues are given by:
\be \widehat{\left[{\frac{{\rm sgn}(p)}{\sqrt{|p|}}}\right]}\,\,
|\L\rangle = \f{6}{\gamma\lp^2}\, (V_{\L+1}^{1/3} -
 V^{1/3}_{\L-1}) \, |\L\rangle\, .\end{equation}
where $V_\L$ is the eigenvalue of the volume operator (see
(\ref{vol})). Next, we note a key property of the triad operator:
It is bounded above! The upper bound is obtained at the value $\L=
1$:
\be
 |p|^{-\frac{1}{2}}_{\rm max}\, =\, \sqrt{\frac{12}{\gamma}}\,
 \lp^{-1}\,.
\end{equation}
This is a striking result because $p$ admits a \emph{normalized}
eigenvector with zero eigenvalue. Since in the classical theory
the curvature is proportional to $p^{-1}$, in quantum theory, it
is bounded above by $(12/\g) \lp^{-2}$. Note that $\hbar$ is
essential for the existence of this upper bound; as $\hbar$ tends
to zero, the bound goes to infinity just as one would expect from
classical considerations. This is rather reminiscent of the
situation with the ground state energy of the hydrogen atom in
non-relativistic quantum mechanics, $E_o = -(m_{\rm e} e^4/2)
(1/\hbar)$, which is bounded from below because $\hbar$ is
non-zero.

In light of this surprising result, let us re-examine the physical
meaning of the quantization procedure. (Using homogeneity and
isotropy, we can naturally convert volume scales in to length
scales. From now on, we will do so freely.) In the classical
Poisson bracket, we replaced the connection coefficient $c$ by the
holonomy along an edge of the elementary cell ${\cal V}$ because
there is no operator on $\H^{S}_{\rm grav}$ corresponding to $c$.
Since the cell has volume $V_o$ \emph{with respect to the fiducial
metric} $\q_{ab}$, the edge has length $V_o^{1/3}$. While this
length can be large, what is relevant is the \emph{physical}
length of this edge and we will now present two arguments showing
that the physical length is of the order of a Planck length. The
first uses states. Let us begin by noting that, being a function
of the connection, (matrix elements of) the holonomy itself
determine a \emph{quantum state} ${\cal N}_{\L = 1}(A) =
e^{i\frac{c}{2}}$. In this state, the physical volume of the cell
${\cal V}$ is not $V_o$ but $(\g/6)^{3/2}\lp^3$. Hence, the
appropriate \emph{physical} edge length is $(\g /6)^{1/2}\lp$,
and this is only of the order of the Planck length.%
\footnote{Black hole entropy calculations imply that we should set
$\g = \f{\ln 2}{\sqrt{3}\pi}$ to recover the standard quantum
field theory in curved space-times from quantum geometry.
\cite{abck}.}
The same conclusion is reached by a second argument based on the
\emph{operator} $\hat{h}_i$: Since $e^{i\frac{c}{2}}|\L\rangle =
|\L +1 \rangle$ for any $\L$, the holonomy operator changes the
volume of the universe by `attaching' edges of physical length
$(\g /6)^{1/2}(|\L +1|^{1/2} -|\L|^{1/2})\,\lp$.%
\footnote{The square-root of $\L$ features rather than $\L$ itself
because $\hat{p}$ corresponds to the \emph{square} of the scale
factor $a$ and we chose to denote its eigenvalues by
$(\g\mu/6)\lp^2$.}
These arguments enable us to interpret the quantization procedure
as follows. There is no direct operator analog of $c$; only
holonomy operators are well-defined. The `fundamental' triad
operator (\ref{qtriad1}) involves holonomies along \emph{Planck
scale} edges. In the classical limit, we can let the edge length
go to zero and then this operator reduces to the classical triad,
the Poisson bracket $\{c,\, V^{{1}/{3}} \}$.

Since the classical triad diverges at the big bang, it is perhaps
not surprising that the `regularization' introduced by quantum
effects ushers-in the Planck scale. However, the mechanism by
which this came about is new and conceptually important. For, we
did not introduce a cut-off or a regulator; the classical
expression (\ref{e2}) of the triad coefficient we began with is
\emph{exact}. Since we did not `regulate' the classical
expression, the issue of removing the regulator does not arise.
Nonetheless, it \emph{is} true that the quantization procedure is
`indirect'. However, this was necessary because \emph{the spectrum
of the momentum operator $\hat{p}$} (or of the `scale factor
operator' corresponding to $a$) \emph{is discrete} in the sense
detailed in section \ref{s3.2}. Had the Hilbert space $\H^{S}_{\rm
grav}$ been a direct integral of the eigenspaces of $\hat{p}$
---rather than a direct sum---  the triad operator could then have
been defined directly using the spectral decomposition of
$\hat{p}$ and would have been unbounded above.

Indeed, this is precisely what happens in geometrodynamics. There,
$p$ and $c$ themselves are elementary variables and the Hilbert
space is taken to be the space of square integrable functions of
$p$ (or, rather, of $a \sim {|p|^{1/2}}$). Then, $p$ has a
genuinely continuous spectrum and its inverse is a self-adjoint
operator, defined in terms the spectral decomposition of $p$ and
is \emph{unbounded} above. By contrast, in loop quantum gravity,
quantization is based on holonomies ---the Wilson lines of the
gravitational connection. We carried this central idea to the
symmetry reduced model. As a direct result, as in the full theory,
we were led to a non-standard Hilbert space $\H^S_{\rm grav}$.
Furthermore, we were led to consider almost periodic functions of
$c$ ---rather than $c$ itself--- as `elementary' and an operator
corresponding to $c$ is not even defined on $\H^S_{\rm grav}$. All
eigenvectors of $p$ are now \emph{normalizable}, including the one
with zero eigenvalue. Hence, to define the triad operator, one
simply can not repeat the procedure of geometrodynamics. We are
led to use the alternate procedure followed above. Of course one
could simply invent a regularization scheme just for this symmetry
reduced model. A key feature of our procedure is that it was not
so invented; it is the direct analog of the procedure followed to
address the same issue in the full theory \cite{tt}.

Finally, let us return to the expression of the quantum operator
(\ref{qtriad1}). Since the fact that it is bounded is surprising,
it is important to verify that the final result has physically
reasonable properties. The first obvious requirement is that,
since the triad coefficient ${\rm sgn}\, p /|p|^{\f{1}{2}}$ is a
function only of $p$, the triad operator should commute with
$\hat{p}$. A priori there is no guarantee that this would be the
case. Indeed, the expression (\ref{qtriad1}) of the triad operator
involves $c$ as well. However, as we saw, this condition is in
fact met. A second non-trivial requirement comes from the fact
that the triad coefficient and the momentum are algebraically
related in the classical theory: $p \cdot ({\rm
sgn}p/{|p|^{1/2}})^2 = 1$. A key criterion of viability of the
triad operator is that this relation should be respected in an
appropriate sense. More precisely, we can tolerate violations of
this condition on states only in the Planck regime; the equality
must be satisfied to an excellent approximation on states with
large $\L$ (i.e., with large volume). Is this the case? We have:
\ba \label{approximate}
 \f{6}{\gamma\lp^2}\, (V_{\L+1}^{1/3} - V^{1/3}_{\L-1}) &=&
 \sqrt{\frac{6|\L|}{\gamma\lp^2}}\,\,\left(\sqrt{|1+ 1/\L|}-
 \sqrt{|1- 1/\L|}\right)\nonumber\\
 &=& {\rm sgn}\L\,\sqrt{\frac{6}{\gamma|\L|\lp^2}}\,\,(1+O(\L^{-2}))
\ea
%
Thus, up to order $O(\L^{-2})$, the eigenvalue of the triad operator
is precisely ${\rm sgn} p_{\L}/\sqrt{|p_\L|}$, where $p_\L$ is the
eigenvalue of $\hat{p}$ (see (\ref{vol})). On states representing a
large universe ($|\L| \gg 1$), the classical algebraic relation
between the triad coefficient and $p$ is indeed preserved to an
excellent approximation. Violations are significant only on the
eigen-subspace of the volume operator with eigenvalues of the order of
$\lp^3$ or less, i.e., in the fully quantum regime.

\emph{Remark}: In (\ref{e2}), we used holonomies along edges of
the elementary cell ${\cal V}$. While this choice is natural
because the cell is needed for classical considerations in any
case, one might imagine using, instead, edges of length $|\mu_o|
V_o^{1/3}$. Had this been done, the replacement of (\ref{e2})
would have again provided an \emph{exact} expression of the triad
coefficient in the classical theory. However, to meet the second
criterion above, one would be forced to choose $|\mu_o| \sim 1$
and we would be back with the `natural' choice made above.
Nonetheless, since there \emph{is} a quantization ambiguity, the
numerical coefficients in the final results (e.g., the precise
value of the upper bound of the triad operator spectrum) should
not be attached direct physical significance. In particular, for
lessons for the full theory, one should use only the qualitative
features and orders of magnitudes. The numerical values can be
arrived at only through a systematic reduction of the full quantum
theory, where the precise value of $|\mu_o| (\sim 1)$ should
emerge, e.g., as the lowest eigenvalue of an appropriate geometric
operator. We will return to this issue in the next section.

\section{Quantum dynamics: The Hamiltonian constraint}
\label{s4}

Since the curvature is bounded above on the entire kinematical
Hilbert space $\H^S_{\rm grav}$, one might expect that the
classical singularity at the big bang would be naturally resolved
in the quantum theory. In this section we will show that this is
indeed the case.

\subsection{The quantum Hamiltonian constraint}
\label{s4.1}

In section \ref{s1}, we reduced the Hamiltonian constraint to
(\ref{scalar}). However, we can not use this form of the
constraint directly because it is cast in terms of the connection
$c$ itself rather than holonomies. One can `regulate' it in terms
of holonomies and then pass to quantum theory. However, to bring
out the close similarity of the regularization procedure with the
one followed in the full theory \cite{tt}, we will obtain the same
expression starting from the expression of the classical
constraint in the \emph{full} theory:
\ba C_{\rm grav} &:=& \int_{\cal V} d^3x N\, e^{-1}
\left(\epsilon_{ijk} F_{ab}^iE^{aj}E^{bk}- 2(1+\gamma^2)
K_{[a}^iK_{b]}^jE_i^aE_j^b \right)\nonumber\\
&=& -\gamma^{-2}\int_{\cal V}d^3x\,  N\, \epsilon_{ijk}F_{ab}^i\,\,
e^{-1} {E^{aj}E^{bk}} \ea
where $e:=\sqrt{|\det E|}\,\, \text{sgn}(\det E)$. We restricted the
integral to our cell ${\cal V}$ (of volume $V_o$ with respect to
$\q_{ab}$) and, in the second step, exploited the fact that for
spatially flat, homogeneous models the two terms in the full
constraint are proportional to each other (one can also treat both
terms as in the full theory without significant changes
\cite{mb7}). Because of homogeneity, we can assume that the lapse $N$
is constant and, for definiteness, from now onwards we will set it to
one.

As a first step in constructing a Hamiltonian constraint operator
we have to express the curvature components $F_{ab}^i$ in terms of
holonomies. We will use the procedure followed in the full theory
\cite{tt} (or in lattice gauge theories). Consider a square
$\a_{ij}$ in the $i$-$j$ plane spanned by two of the triad vectors
$\e^a_i$, each of whose sides has length $\L_o V_o^{1/3}$ with
respect to the fiducial metric $\q_{ab}$.%
\footnote{In a model with non-zero intrinsic curvature, those
edges would not form a closed loop. The issue of how to deal with
intrinsic curvature is discussed in \cite{bdv}.}
Then, `the $ab$ component' of the curvature is given by
\be \label{F} F_{ab}^i\tau_i = \w^i_a\,\, \w^j_b\,\,
\left(\f{h^{(\L_o)}_{\a_{ij}}-1 }{\L_o^2V_o^{2/3}} +
O(c^3\L_o)\right) \end{equation}
The holonomy $h^{(\L_o)}_{\a_{ij}}$ in turn can be expressed as
\be
 h^{(\L_o)}_{\alpha_{ij}}=h_i^{(\L_o)} h_j^{(\L_o)} (h_i^{(\L_o)})^{-1}
 (h_j^{(\L_o)})^{-1}
\end{equation}
where, as before, holonomies along individual edges are given by
\be h_i^{(\L_o)}:=\cos \f{\L_o c}{2} + 2 \sin \f{\L_o c}{2}\,
\tau_i \end{equation}

Next, let us consider the triad term $\epsilon_{ijk}\, e^{-1}
E^{aj} E^{bk}$ in the expression of the Hamiltonian constraint.
Since the triad is allowed to become degenerate, there is a
potential problem with the factor $e^{-1}$. In the reduced model, $e$
vanishes only when the triad itself vanishes and hence the
required term $\epsilon_{ijk}\, e^{-1} E^{aj} E^{bk}$ can be
expressed as a non-singular function of $p$ and the fiducial
triads. In the full theory, the situation is more complicated and
such a direct approach is not available. There is nonetheless a
procedure to handle this apparently singular function \cite{tt}:
one expresses it as a Poisson bracket between holonomies and the
volume function as in section \ref{s3.3} and then promotes the
resulting expression to an operator. To gain insight in to this
strategy, here we will follow the same procedure. Thus, let us
begin with the identity on the symmetry reduced phase space
$\Gamma^S_{\rm grav}$:
\be \label{cotriad}
 \epsilon_{ijk}\tau^i\, e^{-1}E^{aj}E^{bk}\,
=\, -2 (8\pi \gamma G \L_o V_o^{1/3})^{-1} \, \epsilon^{abc}\, \w^k_c\,
 h_k^{(\L_o)}\,
\{h_k^{(\L_o)}{}^{-1}, V\} \end{equation}
where $h_k^{(\L_0)}$ is the holonomy along the edge parallel to
the $k$th basis vector of length $\L_o V_o^{1/3}$ with respect to
$\q_{ab}$. Note that, unlike the expression (\ref{F}) for
$F_{ab}^i$, (\ref{cotriad}) is exact, i.e. does not depend on the
choice of $\L_o$.

Collecting terms, we can now express the gravitational part of
the constraint as:
\ba \label{reg}
 C_{\rm grav} &\!\!\!\! = \!\!\!\!& -4(8\pi\gamma^3\L_o^3 G)^{-1} \sum_{ijk}
\epsilon^{ijk}{\rm tr}(h_i^{(\L_o)}h_j^{(\L_o)}
h_i^{(\L_o)-1}h_j^{(\L_o)-1} h_k^{(\L_o)}\{h_k^{(\L_o)-1},V\})
\nonumber\\
&&+ O({c^3\L_o}) \ea
where, the term proportional to identity in the leading
contribution to $F_{ab}^i$ in (\ref{F}) drops out because of the
trace operation and where we used
$\epsilon^{abc}\,\w^i_a\w^j_b\w^k_c= \sqrt{\q}\,\epsilon^{ijk}$.
Note that, in contrast to the situation with triads in section
\ref{s3.3}, now the dependence on $\L_o$ does \emph{not} drop out.
However, one can take the limit $\L_o \rightarrow 0$. Using the
explicit form of the holonomies $h^{(\L_o)}_i$, one can verify
that the leading term in (\ref{reg}) has a well-defined limit
which equals \emph{precisely} the classical constraint. Thus, now
$\L_o$ ---or the length of the edge used while expressing $F_{ab}$
in terms of the holonomy around the square $\alpha_{ij}$--- plays
the role of a regulator. Because of the presence of the curvature
term, there is no natural way to express the constraint
\emph{exactly} in terms of our elementary variables; a limiting
procedure is essential. This faithfully mirrors the situation in
the full theory: there, again, the curvature term is recovered by
introducing small loops at vertices of graphs and the classical
expression of the constraint is recovered only in the limit in
which the loop shrinks to zero.

Let us focus on the leading term in (\ref{reg}). As in the full
theory, this term is manifestly finite and can be promoted to a
quantum operator directly. The resulting regulated constraint is:
\ba \label{qreg}
 \hat{C}_{\rm grav}^{(\L_o)} &=& 4i(\gamma^3\L_o^3\lp^2)^{-1}
 \sum_{ijk}\epsilon^{ijk}{\rm tr}(\hat{h}_i^{(\L_o)}
 \hat{h}_j^{(\L_o)} \hat{h}_i^{(\L_o)-1}\hat{h}_j^{(\L_o)-1}
  \hat{h}_k^{(\L_o)}[\hat{h}_k^{(\L_o)-1},\hat{V}])\nonumber\\
  &=& 96i(\gamma^3\L_o^3\lp^2)^{-1} \sin^2\frac{\L_oc}{2}
  \cos^2\frac{\L_oc}{2} \\
  &&\times \left(\sin\frac{\L_oc}{2}\hat{V}
  \cos\frac{\L_oc}{2}- \cos\frac{\L_oc}{2}\hat{V}
  \sin \frac{\L_oc}{2}\right)\nonumber
\ea
Its action on the eigenstates of $\hat{p}$ is
\be
 \hat{C}_{\rm grav}^{(\L_o)}\, |\L\rangle=3(\gamma^3\L_o^3\lp^2)^{-1}
 (V_{\L+\L_o}-V_{\L-\L_o}) (|\L+4\L_o\rangle-
 2|\L\rangle+ |\L-4\L_o\rangle)\, .
\end{equation}
On physical states, this action must equal that of the matter
Hamiltonian $-8\pi G\hat{C}_{\rm matter}$.

Now, the limit $\L_o \rightarrow 0$ of the classical expression
(\ref{reg}) exists and equals the classical Hamiltonian constraint
which, however, contains $c^2$ (see (\ref{scalar})). Consequently,
the naive limit of the operator $\hat{C}_{\rm grav}^{(\L_o)}$ also
contains $\hat{c}^2$. However, since $\hat{c}^2$ is not
well-defined on $\H^S_{\rm grav}$, now the limit as $\L_o
\rightarrow 0$ fails to exist. Thus, we can not remove the
regulator in the quantum theory of the reduced model. In the full
theory, by contrast, one \emph{can} remove the regulator and
obtain a well-defined  action on \emph{diffeomorphism invariant}
states \cite{tt}. This difference can be directly traced back to
the assumption of homogeneity.%
\footnote{In the full theory, one triangulates the manifold with
tetrahedra of coordinate volume $\L_o^3$  and writes the integral
$C(N):= \int d^3x N\, \epsilon_{ijk} F_{ab}^i\, e^{-1}
E^{aj}E^{bk}$ as the limit of a Riemann sum,  $C(N) = \lim_{\L_o
\to 0}\, \sum \L_o^3 N \epsilon_{ijk} F_{ab}^i\, e^{-1}
E^{aj}E^{bk}$, where the sum is over tetrahedra. If we now replace
$F$ by a holonomy around a square $\alpha$ of length $\L_o$, $F \,
\sim \, \L_o^{-2}\, (h_\alpha -1)$, and the triad term by a
Poisson bracket, $e^{-1} E E\, \sim \, \L_o^{-1}\, {h} \{ h^{-1},
\, V \}$, and pass to quantum operators, we obtain $\hat{C}(N) \,
\sim\, \lim \sum {\rm tr} \hat{h}_\alpha \hat{h} [\hat{h}^{-1},\,
\hat{V}]$. The $\L_o$ factors cancel out but, in the sum, the
number of terms goes to infinity as $\L_o \rightarrow 0$. However,
the action of the operator on a state based on any graph is
non-trivial \emph{only} at the vertices of the graph whence only a
finite number of terms in the sum survive and these have a well
defined limit on diffeomorphism invariant states. In the reduced
model, because of homogeneity, all terms in the sum contribute
equally and hence the sum diverges.}
In the full theory, there is nonetheless a quantization ambiguity
associated with the choice of the $j$ label used on the new edges
introduced to define the operator corresponding to $F_{ab}$
\cite{gr}. That is, in the full theory, the quantization procedure
involves a pair of labels $(e,j)$ where $e$ is a continuous label
denoting the new edge and $j$ is a discrete label denoting the
spin on that edge. Diffeomorphism invariance ensures that the
quantum constraint is insensitive to the choice of $e$ but the
dependence on $j$ remains as a quantization ambiguity. In the
reduced model, diffeomorphism invariance is lost and the pair $(e,
j)$ of the full theory collapses into a single continuous label
$\L_o$ denoting the length of the edge introduced to define
$F_{ab}$. The dependence on $\L_o$ persists ---there is again a
quantization ambiguity but it is now labelled by a
\emph{continuous} label $\L_o$. Thus, comparison of the situation
with that in the full theory suggests that we should not regard
$\hat{C}_{\rm grav}^{(\L_o)}$ as an approximate quantum
constraint; it is more appropriate to think of the
$\L_o$-dependence in (\ref{qreg}) as a \emph{quantization
ambiguity} in the exact quantum constraint. This is the viewpoint
adopted in loop quantum cosmology.

If one works in the strict confines of the reduced model, there
does not appear to exist a natural way of removing this ambiguity.
In the full theory, on the other hand, one can fix the ambiguity
by assigning the \emph{lowest} non-trivial $j$ value, $j= 1/2$, to
each extra loop introduced to determine the operator analog of
$F_{ab}$. This procedure can be motivated by the following
heuristics. In the classical theory, we could use a loop enclosing
an arbitrarily small area in the $a$-$b$ plane to determine
$F_{ab}$ locally. In quantum geometry, on the other hand, the area
operator (of an open surface) has a lowest eigenvalue $a_o =
(\sqrt{3}\gamma)/4\, \lp^2$ \cite{al5,area} suggesting that it is
\emph{physically} inappropriate to try to localize $F_{ab}$ on
arbitrarily small surfaces. The best one could do is to consider a
loop spanning an area $a_o$, consider the holonomy around the loop
to determine the integral of $F_{ab}$ on a surface of area $a_o$,
and then extract an effective, local $F_{ab}$ by setting the
integral equal to $a_o F_{ab}$. It appears natural to use the same
physical considerations to remove the quantization ambiguity also
in the reduced model. Then, we are led to set the area of the
smallest square spanned by $\alpha_{ij}$ to $a_o$, i.e. to set
$(\gamma \L_o)\, \lp^2 = a_o$, or $\L_o = \sqrt{3}/4$. Thus, while
in the reduced model itself, area eigenvalues can assume
arbitrarily small values, if we `import' from the full theory the
value of the smallest non-zero area eigenvalue, we are naturally
led to set $\L_o = \sqrt{3}/4$. We will do so.

To summarize, in loop quantum cosmology, we adopt the viewpoint
that (\ref{qreg}), with $\L_o = \sqrt{3}/4$, is the `fundamental'
Hamiltonian constraint operator which `correctly' incorporates the
underlying discreteness of quantum geometry and the classical
expression (\ref{scalar}) is an approximation which is valid only
in regimes where this discreteness can be ignored and the
continuum picture is valid. We will justify this assertion in
section \ref{s4.3}.

\subsection{Physical states}
\label{s4.2}

Let us now solve the quantum constraint and obtain physical
states. For simplicity, we assume that the matter is only
minimally coupled to gravity (i.e., there are no curvature
couplings). As in general non-trivially constrained systems, one
expects that the physical states would fail to be normalizable in
the kinematical Hilbert space $\H^S = \H^S_{\rm grav}\otimes
\H^S_{\rm matter}$ (see, e.g., \cite{dm,almmt}). However, as in
the full theory, they do have a natural `home'. We again have a
triplet
$$ \cyl_S \subset \H^S \subset \cyl^\star_S $$
of spaces and physical states will belong to $\cyl_S^\star$, the
algebraic dual of $\cyl_S$. Since elements of $\cyl_S^\star$ need
not be normalizable, we will denote them by $(\Psi|$. (The usual,
normalizable bras will be denoted by $\langle \Psi|$.)

It is convenient to exploit the existence of a triad
representation. Then, every element $(\Psi|$ of $\cyl^\star_S$ can
be expanded as
\be (\Psi| \, = \, \sum_{\L}\, \psi(\phi, \L) \langle\L|\end{equation}
where $\phi$ denotes the matter field and $\langle \L|$ are the
(normalized) eigenbras of $\hat{p}$. Note that the sum is over a
continuous variable $\L$ whence $(\Psi|$ need not be normalizable.
Now, the constraint equation
\be (\Psi|\,\left(\hat{C}_{\rm grav}^{(\L_o)}\, +\,  8\pi
G\hat{C}_{\rm matter}^{(\L_o)} \right)^{\dagger}\, =0\end{equation}
turns into the equation
\ba \label{de1}
 &&(V_{\L+5\L_o}-V_{\L+3\L_o})\psi(\phi, {\L+4\L_o}) -
 2(V_{\L+\L_o}-V_{\L-\L_o})\psi(\phi, \L) \\
 &+&(V_{\L-3\L_o}-V_{\L-5\L_o})\psi (\phi, \L-4\L_o)
 = -\frac{1}{3}\,8\pi G\gamma^3\L_o^3\lp^2\,\,
 \hat{C}_{\rm matter}^{(\L_o)}(\L) \psi(\phi, \L)\nonumber
\ea
for the coefficients $\psi(\phi, \L)$, where $\hat{C}_{\rm
matter}^{(\L_o)}(\L)$ only acts on the matter fields (and depends
on $\L$ via metric components in the matter Hamiltonian). Note
that, even though $\L$ is a continuous variable, the quantum
constraint is a \emph{difference} equation rather than a
differential equation. Strictly, (\ref{de1}) just constrains the
coefficients $\psi(\phi, \L)$ to ensure that $(\Psi|$ is a
physical state. However, since each $\langle \L|$ is an eigenbra
of the volume operator, it tells us how the matter wave function
is correlated with volume, i.e., geometry. Now, if one wishes, one
can regard $p$ as providing a heuristic `notion of time', and then
think of (\ref{de1}) as an evolution equation for the quantum
state of matter with respect to this time. (Note that $p$ goes
from $-\infty$ to $\infty$, negative values corresponding to
triads which are oppositely oriented to the fiducial one. The
classical big-bang corresponds to $p=0$.) While this heuristic
interpretation often provides physical intuition for (\ref{de1})
and its consequences, it is \emph{not} essential for what follows;
one can forego this interpretation entirely and regard (\ref{de1})
only as a constraint equation.

What is the fate of the classical singularity? At the big bang, the
scale factor goes to zero. Hence it corresponds to the state $|\L
=0\rangle$ in $\H^S_{\rm grav}$. So, the key question is whether the
quantum `evolution' breaks down at $\L= 0$. Now, the discrete
`evolution equation' (\ref{de1}) is essentially the same as that
considered in the first papers on isotropic loop quantum cosmology
\cite{mb2,mb3} and that discussion implies that the quantum physics
does \emph{not} stop at the big-bang.

For completeness, we now recall the main argument. The basic idea
is to explore the key consequences of the difference equation
(\ref{de1}) which determine what happens at the initial
singularity. Starting at $\L = -4N \L_o$ for some large positive
$N$, and fixing $\psi(\phi, -4N\L_o)$ and $\psi(\phi,
(-4N+4)\L_o)$, one can use the equation to determine the
coefficients $\psi(\phi, (-4N+4n)\L_0)$ for all $n > 1$,
\emph{provided} the coefficient of the highest order term in
(\ref{de1}) continues to remain non-zero. Now, it is easy to
verify that the coefficient vanishes if and only if $n= N$. Thus,
the coefficient $\psi(\phi, \L\! =\! 0)$ remains undetermined. In
its place, we just obtain a consistency condition constraining the
coefficients $\psi(\phi, \L\!=\! -4)$ and $\psi(\phi, \L\!=\!
-8)$. Now, since  $\psi(\phi, \L\!=\! 0)$ remains undetermined, at
first sight, it may appear that we can not `evolve' past the
singularity, i.e. that the quantum evolution also breaks down at
the big-bang. However, the main point is that this is \emph{not}
the case. For, the coefficient $\psi(\phi, \L\! =\!0)$ just
decouples from the rest. This comes about because, as a detailed
examination shows, the minimally coupled matter Hamiltonians
annihilate $\psi (\phi, \L)$ for $\L=0$ \cite{mb1,mb7} and
$V_{\L_o} = V_{-\L_o}$. Thus, unlike in the classical theory,
evolution does not stop at the singularity; the difference
equation (\ref{de1}) lets us `evolve' right through it. In this
analysis, we started at $\L= -4N\L_o$ because we wanted to test
what happens if one encounters the singularity `head on'. If one
begins at a generic $\L$, the `discrete evolution' determined by
(\ref{de1}) just `jumps' over the classical singularity without
encountering any subtleties.

To summarize, two factors were key to the resolution of the big
bang singularity: i) as a direct consequence of quantum geometry,
the Hamiltonian constraint is now a difference equation rather
than a differential equation as in geometrodynamics; and ii) the
coefficients in the difference equation are such that one can
evolve unambiguously `through' the singularity even though the
coefficient $\psi(\phi, \L=0)$ is undetermined. Both these
features are robust: they are insensitive to factor ordering
ambiguities and persist in more complicated cosmological models
\cite{mb5,mb6}.

Next, let us consider the space of solutions. An examination of
the classical degrees of freedom suggests that the freedom in
physical quantum states should correspond to two functions
\emph{just} of matter fields $\phi$. The space of solutions to the
Hamiltonian constraint, on the other hand is much larger: there
are as many solutions as there are functions $\psi(\phi, \L)$ on
an interval $[\L^\prime- 4\L_o, \L^\prime+ 4\L_o)$, where
$\L^\prime$ is any fixed number. This suggests that a large number
of these solutions may be redundant. Indeed, to complete the
quantization procedure, one needs to introduce an appropriate
inner product on the space of solutions to the Hamiltonian
constraint. The physical Hilbert space is then spanned by just
those solutions to the quantum constraint which have finite norm.
In simple examples one generally finds that, while the space of
solutions to all constraints can be very large, the requirement of
finiteness of norm suffices to produce a Hilbert space of the
physically expected size.

For the reduced system considered here, we have a quantum mechanical
system with a single constraint in quantum cosmology. Hence it should
be possible to extract physical states using the group averaging
technique of the `refined algebraic quantization framework'
\cite{dm,almmt,tt1}.  However, this analysis is yet to be carried out
explicitly and therefore we do not yet have a good control on how
large the physical Hilbert space really is. This issue is being
investigated.

The Hamiltonian constraint equation differs markedly from the
Wheeler-DeWitt equation of geometrodynamics in the Planck regime
because it crucially exploits the discreteness underlying quantum
geometry. But one might expect that in the continuum limit $\mu_o
\rightarrow 0$ ---which, from the quantum geometry perspective, is
physically fictitious but nonetheless mathematically
interesting--- the present quantum constraint equation would
reduce to the Wheeler-DeWitt equation. We will conclude this
sub-section by showing that this expectation is indeed correct in
a precise sense.

To facilitate this comparison, it is convenient to introduce some
notation. Let us set
$$ p = \f{1}{6} \g \L \lp^2 \, .$$
Then, the Wheeler-DeWitt equation can be written as
\be \label{wdw}  \hat{C}_{\rm grav}^{\rm wdw} \psi (\phi, p) :=
\f{2}{3} \lp^4\, [\sqrt{|p|}\, \psi(\phi, p)\, ]'' = 8\pi G \,
\hat{C}_{\rm matter}(p) \psi(\phi, p)\, ,\end{equation}
where the prime denotes derivative with respect to $p$. If we now
further set
$$\tilde{\psi}(p):=\frac{1}{6}\,p_o^{-1}
(V_{6(p+p_o)/\gamma\lp^2}- V_{6(p-p_o)/\gamma\lp^2}) \psi(p)\, ,$$
with $p_o=\gamma\lp^2\L_o/6$, our quantum constraint (\ref{de1})
reduces to:
\ba \label{de2} {\hat{\tilde{C}}}_{\rm grav}^{(\L_o)} \,\,
\tilde\psi(\phi, p) &:=& -\frac{1}{12}\,\lp^4p_o^{-2}\,(\tilde{\psi}(\phi,
p+4p_o) -2\tilde{\psi}(\phi, p)+\tilde{\psi}(\phi, p-4p_o))
\nonumber\\
&=& 8\pi G\hat{C}_{\rm matter}(p)\psi(\phi, p)\, . \ea
From now on we will consider only those `wave functions'
$\tilde\psi(\phi, p)$ which are smooth (more precisely, $C^4$) in
their $p$ dependence. Then, it follows that
\ba \hat{C}_{\rm grav}^{\rm wdw}\, \psi (\phi, p)\, &=&\,
{\hat{\tilde{C}}}_{\rm grav}^{(\L_o)}\,\, \tilde\psi (\phi,p)  +
\lp^4\, O(p_o^2)\, \tilde \psi^{\prime \prime \prime\prime}(\phi,
p) \\
&+& \lp^4\, O(\f{p_o^2}{p^2})\, \tilde\psi^{\prime\prime} (\phi,
p)+ \lp^4\, O(\f{p_o^2}{p^3}) \tilde\psi^\prime (\phi, p) +
\lp^4\, O(\f{p_o^2}{p^4}) \tilde\psi(\phi, p) \nonumber\ea
Hence, in the limit $p_o \rightarrow 0$ (i.e., $\L_o \rightarrow
0$), we have
\be {\hat{\tilde{C}}}_{\rm grav}^{(\L_o)}\,\, \tilde\psi(\phi, p)
\,\, \mapsto \,\, \hat{C}_{\rm grav}^{\rm wdw} \psi (\phi, p)\,
\end{equation}
whence the discrete equation (\ref{de2}) reduces \emph{precisely}
to the Wheeler-DeWitt equation (\ref{wdw}). Put differently, it
has turned out that (\ref{de2}) is a well-defined discretization
of (\ref{wdw}).

One can also ask a related but distinct question: Is there a sense
in which solutions to the Wheeler-DeWitt equation are approximate
solutions to the `fundamental' discrete evolution equation? The
answer is again in the affirmative. Let us restrict ourselves to
the part of the $p$-line where $p \gg p_o$, i.e., where the
quantum volume of the universe is very large compared to the
Planck scale. Consider the restriction, to this region, of a
smooth solution $\psi (\phi, p)$ to (\ref{wdw}) and assume that it
is slowly varying at the Planck scale in the sense that
$\tilde\psi/\tilde\psi^\prime\, \sim s$,
$\tilde\psi/\tilde\psi^{\prime\prime}\, \sim s^2$, etc, with $p_o
\ll s \le p$. Then, $\tilde{\psi}(\phi,p)$ is an approximate
solution to the `fundamental' quantum constraint (\ref{de2}) in
the sense that:
 \ba &&\left[ 1 +
   O(\f{p_o^2}{s^2})+ O(\f{p_o^2}{p^2})+ O(\f{p_o^2s}{p^3})+
   O(\f{p_o^2s^2}{p^4})
   \right]\,  \hat{\tilde{C}}_{\rm
grav}^{(\L_o)} \tilde\psi (\phi, p)\\
&& \quad = \, 8\pi G \hat{C}_{\rm matter}(p)\psi(\phi, p)\, . \nonumber
\ea
Note that, in contrast to the discussion about the relation between
the two \emph{equations}, we can not take the limit $p_o \rightarrow
0$ because we are now interested in the discrete evolution. The
solution to the Wheeler-DeWitt equation is an approximate solution to
the fundamental equation only to the extent that terms of the order
$O(p_o^2/s^2),\,\, O({p_o^2}/{p^2}),\,\, O({p_o^2s}/{p^3}),\,\,
O({p_o^2s^2}/{p^4})$ are negligible.

We will conclude with three remarks.

1) We saw in section \ref{s4.1} that the $\L_o \rightarrow 0$
limit of the quantum constraint operator $\hat{C}^{(\L_o)}_{\rm
grav}$ does not exist on $\H^S_{\rm grav}$. Yet, in the above
discussion of the `continuum limit', we were able to take this
limit. The resolution of this apparent paradox is that the limit
is taken on a certain sub-space of $\cyl^\star$, consisting of
\emph{smooth} functions of $p$ and none of these states belong to
$\H^S_{\rm grav}$. Indeed, since elements of $\H^S_{\rm grav}$
have to be normalizable with respect to the inner product
(\ref{ip}), they can have support only on a \emph{countable}
number of points; they cannot even be continuous. In particular,
solutions to the Wheeler-DeWitt equation can not lie in $\H^S$;
they belong only to the enlargement $\cyl^\star$ of $\H^S$.

2) There is a close mathematical similarity between quantum
cosmology discussed here and the `polymer particle' example
discussed in \cite{afw}. In that example, following the loop
quantum gravity program, a new representation of the Weyl algebra
is introduced for a point particle in non-relativistic quantum
mechanics. In this representation, the Weyl operators are
unitarily implemented but weak continuity, assumed in the Von
Neumann uniqueness theorem, is violated for one of the two
1-parameter unitary groups. As a result (although the position
operator exists) the momentum operator ---the generator of
infinitesimal space translations--- fails to exist. This is meant
to reflect the underlying discreteness of geometry. The
`fundamental' quantum evolution is given by a difference equation.
But there is a precise sense in which the standard Schr\"odinger
evolution is recovered in the regime of validity of
non-relativistic quantum mechanics. The `fundamental' discrete
evolution is analogous to the present `fundamental' quantum
constraint (\ref{de2}) while the Schr\"odinger equation is the
analog of the Wheeler-DeWitt equation (\ref{wdw}). Therefore,
details of the polymer particle analysis provide good intuition
for the `mechanism' that allows loop quantum cosmology to be very
different from the standard one in the Planck regime and yet agree
with it when the universe is large compared to the Planck scale.

3) As mentioned in section \ref{s1}, in this paper we do not
address the difficult issue of systematically \emph{deriving}
quantum cosmology from full loop quantum gravity. Indeed, since
$\cyl_S \not\subset \cyl$, at first it seems it would be difficult
to relate the two theories. However, note that the physical states
of the symmetry reduced model are elements of $\cyl_S^\star$ while
those of the full theory are elements of $\cyl^\star$, and
$\cyl_S^\star $ \emph{is} contained in $\cyl^\star$: elements of
$\cyl_S^\star$ are those distributions on the full quantum
configuration space $\Ab$ which are supported only on the subspace
$\Ab_S$ of symmetric connections \cite{bk}. In particular,
solutions to the quantum constraint discussed in this section do
belong to $\cyl^\star$. Therefore, it should be possible to
recover such states by first considering the full quantum theory
and then carrying out a symmetry reduction.

\subsection{Classical limit}
\label{s4.3}

In section \ref{s4.1}, we found that the gravitational part of the
Hamiltonian constraint could not be introduced by a
straightforward `quantization' of the classical constraint
(\ref{scalar}) because there is no direct operator analog of $c$
on $\H^S_{\rm grav}$. We then followed the strategy adopted in the
full theory to arrive at the expression (\ref{qreg}) of
$\hat{C}^{(\L_o)}_{\rm grav}$. To ensure that this is a viable
quantization, we need to show that (\ref{qreg}) does reduce to
(\ref{scalar}) in the classical limit. In this sub-section, we
will carry out this task.

For this purpose we will use coherent states peaked at points
$(c_o, N\L_o)$ of the classical phase space where $N\gg 1$ (i.e.
the volume of the universe is very large compared to the Planck
volume) and $c_o \ll 1$ (late times, when the extrinsic curvature
is small compared to the fiducial scale $V_o^{-1/3}$). At such
large volumes and late times, one would expect quantum corrections
to become negligible. The question then is whether the expectation
value of the quantum constraint $\hat{C}^{(\L_o)}_{\rm grav}$ in
these coherent states equals the classical constraint
(\ref{scalar}) modulo negligible corrections. If so,
$\hat{C}^{(\L_o)}_{\rm grav}$ would have the correct classical
limit.

To construct a coherent state, we also have to specify the width
$d$ of the Gaussian (i.e., `tolerance' for quantum fluctuations of
$p$). Now, since the quantum fluctuations in the volume of the
universe must be much smaller than the volume itself, $d \ll
N\L_o$ and since we also want the uncertainty in $c$ to be small,
we must have $\L_o \ll d$. A coherent state of the desired type is
then given by:
\be \label{coh}|\Psi\rangle \,=\,
\sum_{n}\,\left[\exp\,(-((n-N)^2\f{\L_o^2}{2d^2})\, \,\exp\,(-i
((n-N)\L_o)\f{c_o}{2})\right]\,\, |n\L_o\rangle \end{equation}
(More precisely, $|\Psi\rangle$ is the `shadow' on the regular
lattice $\L=n\L_o$ of the coherent state in $\cyl^\star$ uniquely
selected by the triplet $(c_o, N\L_o, d)$. For details, see
\cite{afw}, Section 4.) Our task is to compute the expectation
value
\be\label{exp}
 \langle\hat{C}_{\rm grav}^{(\L_o)}\rangle =
 \frac{\langle\Psi|\hat{C}_{\rm grav}^{(\L_o)}
 |\Psi\rangle}{\langle\Psi|\Psi\rangle}
\end{equation}
of the constraint operator (\ref{de2}):
\[
 \hat{C}_{\rm grav}^{(\L_o)} |\L\rangle=3(\gamma^3\L_o^3\lp^2)^{-1}
 (V_{\L+\L_o}-V_{\L-\L_o}) \,(|\L+4\L_o\rangle-
 2|\L\rangle+ |\L-4\L_o\rangle)\,.
\]

Let us first calculate the expectation value. Setting $\epsilon
:= \L_o/d$, we have:
\begin{eqnarray*}
 \langle\Psi|\hat{C}|\Psi\rangle &=&
 \sum_{n,n'}\, \exp(-\frac{1}{2}\epsilon^2((n'-N)^2+(n-N)^2))
 e^{i\f{c_o}{2}(n'-n)\L_o}\, \langle n'\L_o|\hat{C}|n\L_o\rangle\\
 &=& 3(\gamma^3\L_o^3\lp^2)^{-1}
 \sum_{n,n'}\exp(-\frac{1}{2}\epsilon^2((n'-N)^2+(n-N)^2))
 e^{i\f{c_o}{2}(n'-n)\L_o}\\
 &&\times (V_{(n+1)\L_o}-V_{(n-1)\L_o})\, \langle
 n'\L_o| (|(n+4)\L_o\rangle-2|n\L_o\rangle+ |(n-4)\L_o\rangle)\\
 &=& 3(\gamma^3\L_o^3\lp^2)^{-1} \Bigl[ e^{2ic_o\L_o} \sum_n
 \exp(-\frac{1}{2}\epsilon^2((n+4-N)^2+ (n-N)^2))\\
 &&\qquad\qquad\times(V_{(n+1)\L_o}-V_{(n-1)\L_o})\\
 && -2\sum_n \exp(-\epsilon^2(n-N)^2)\,
 (V_{(n+1)\L_o}-V_{(n-1)\L_o})\\
 &&+ e^{-2ic_o\L_o}\sum_n
 \exp(-\frac{1}{2}\epsilon^2((n-4-N)^2+ (n-N)^2))\\
 &&\qquad\qquad \times(V_{(n+1)\L_o}-V_{(n-1)\L_o})\Bigr]\,.
\end{eqnarray*}
To simplify this expression further, we note that all three sums
in this expression are of the same form and focus on the first:
\begin{eqnarray*}
 \sum_n\, \exp(-\frac{1}{2}\epsilon^2((n+4-N)^2 &+& (n-N)^2))\,
 (V_{(n+1)\L_o}-V_{(n-1)\L_o})\nonumber\\ &=&
 e^{-4\epsilon^2}\sum_n\, e^{-\epsilon^2(n-N)^2}\,
 (V_{(n-1)\L_o}-V_{(n-3)\L_o})
\end{eqnarray*}
where we have completed the square in the exponential and shifted
the summation index by $2$. To compute this sum, as in \cite{afw},
we use the Poisson resummation formula
\be \label{poisson}
 \sum_n\, e^{-\epsilon^2(n-N)^2}f(n)\, =\, \sum_n\, \int
 e^{-\epsilon^2(y-N)^2}f(y)\, e^{2\pi iyn}\, dy\,.
\end{equation} This integral can be evaluated using the steepest descent
approximation (see Appendix). One obtains:
\begin{eqnarray}
 \sum_n\, e^{-\epsilon^2(n-N)^2}\, f(n)
 &=&\f{\sqrt{\pi}}{\epsilon}\, \sum_n\, f(N+\f{i\pi n}{\epsilon^2})
\, e^{-\f{\pi^2n^2}{\epsilon^2}+ 2\pi inN}\, (1+
O((N\epsilon)^{-2}))\nonumber\\
 &=&\f{\sqrt{\pi}}{\epsilon} f(N)(1+ O(e^{-\pi^2/\epsilon^2})+
 O((N\epsilon)^{-2}))
\end{eqnarray}
where, in the last step, we used the fact that, since $\epsilon\ll
1$, terms with $n\not=0$ are suppressed by the exponential. (Note
that $N\epsilon \gg 1$ because $N\L_o \gg d$, i.e., the
permissible quantum fluctuation in the volume of the universe is
much smaller than the volume of the universe at the phase space
point under consideration.)

Finally, we can collect terms to compute the expectation value
(\ref{exp}). Using $\langle\Psi|\Psi\rangle= (\sqrt{\pi}
/\epsilon) (1+O(e^{-\pi^2/\epsilon^2}))$, we have
\begin{eqnarray}
 \langle\hat{C}\rangle_S &=&
 3(\gamma^3\L_o^3\lp^2)^{-1}\left[e^{-4\epsilon^2} e^{2ic_o\L_o}
 (V_{(N-1)\L_o}-V_{(N-3)\L_o})\right. \nonumber\\ 
 &&-2 (V_{(N+1)\L_o}-V_{(N-1)\L_o})
  + \left.e^{-4\epsilon^2} e^{-2ic_o\L_o}
 (V_{(N+3)\L_o}-V_{(N+1)\L_o})\right] \nonumber\\
 &&\times (1+\, O(e^{-{\pi^2}/{\epsilon^2}})+
 O((N\epsilon)^{-2}))\nonumber\\
 &=& \frac{1}{2} (\gamma^2\L_o^2)^{-1} \sqrt{\gamma\L_o\lp^2/6}
 \left[e^{-4\epsilon^2} e^{2ic_o\L_o}
 ((N-1)^{\frac{3}{2}}-(N-3)^{\frac{3}{2}})\right.\nonumber\\
 && -\left.2((N+1)^{\frac{3}{2}}-(N-1)^{\frac{3}{2}}) +
 e^{-4\epsilon^2} e^{-2ic_o\L_o}
 ((N+3)^{\frac{3}{2}}-(N+1)^{\frac{3}{2}})\right]\nonumber\\
 && \times(1+O(e^{-\pi^2/\epsilon^2})+ O((N\epsilon)^{-2}))\nonumber\\
 &=& \frac{3}{2} (\gamma^2\L_o^2)^{-1} \sqrt{\gamma\L_o\lp^2N/6} \,
 (e^{-4\epsilon^2} e^{2ic_o\L_o} -2+ e^{-4\epsilon^2}
 e^{-2ic_o\L_o})\nonumber\\
 &&\times(1+ O(e^{-\pi^2/\epsilon^2})+
 O((N\epsilon)^{-2})+ O(N^{-1}))\nonumber\\
 &=&-6\gamma^{-2}c_o^2 \sqrt{P}(1+
 O((N\epsilon)^{-2})+ O(N^{-1})+O(\epsilon^2)+ O(c_o^2))
\end{eqnarray}
where we have set $P:=\frac{1}{6}\gamma\L_o\lp^2N$ and used the fact
that $c_o\ll 1$ and $\L_o \sim 1$ (we also dropped corrections of
order $e^{-\pi^2/\epsilon^2}$ since they are always dwarfed by those
of order $\epsilon^2$). Thus, the expectation value equals the
classical constraint (\ref{scalar}) up to small corrections of order
$c_o^2$, $\lp^4/(P\epsilon)^2$, $\lp^2/P$ and $\epsilon^2$. (Note that
each of them can dominate the other corrections depending on the
values of the different parameters.) Hence, (\ref{qreg}) is a viable
quantization of the classical expression (\ref{scalar}).

\section{Discussion}
\label{s5}

Let us begin with a brief summary of the main results. In section
\ref{s2}, we carried out a systematic symmetry reduction of the
phase space of full general relativity in the connection
variables. In the spatially flat model considered here, our
connection coefficient $c$ equals the only non-trivial (i.e.
dynamical) component of the extrinsic curvature (modulo a factor
of $\g$) and our conjugate momentum $p$ equals the only
non-trivial metric component (modulo ${\rm sgn}\, p$). Hence, our
symmetry reduced Hamiltonian description is the same as that of
geometrodynamics. By contrast, we saw in section \ref{s3} that
quantum theories are dramatically different \emph{already at the
kinematic level}. In loop quantum cosmology, the Hilbert space
$\H^S_{\rm grav}$ is spanned by almost period functions of $c$
while in geometrodynamics it would be spanned by square-integrable
functions of $c$. The intersection between the two Hilbert spaces
is only the zero element! In loop quantum cosmology, the
fundamental operators are $\hat{p}$ and $\hat{\cal N}_\L =
\widehat{\exp (i\L c/2)}$; unlike in geometrodynamics, there is no
operator corresponding to $c$ itself. Although $\hat{p}$ is
unbounded and its spectrum consists of the entire real line,
\emph{all its eigenvectors are normalizable} and the Hilbert space
is the direct \emph{sum} of the 1-dimensional sub-spaces spanned
by the eigenspaces. In geometrodynamics, on the other hand, none
of the eigenvectors of $\hat{p}$ is normalizable; the Hilbert
space is a direct \emph{integral} of its `eigenspaces'. This
marked difference is responsible for the fact that, while the
triad operator (which encodes the inverse of the scale factor) is
unbounded in geometrodynamics, it is \emph{bounded} in loop
quantum cosmology. Consequently, in the state corresponding to the
classical singularity, the curvature is large, but it does not
diverge in loop quantum cosmology.

In section \ref{s4} we discussed the Hamiltonian constraint, i.e.,
quantum dynamics. Because there is no direct operator analog of
$c$, we had to introduce the constraint operator
$\hat{C}^{(\L_o)}_{\rm grav}$ by an indirect construction. Here,
we followed the strategy used in the full theory \cite{tt},
expressing curvature $F_{ab}^i$ of the gravitational connection
$A_a^i$ in terms of its holonomies around suitable loops. In the
full theory, one can take the limit as the loop shrinks to zero
and obtain a well-defined operator on diffeomorphism invariant
states. The reduced model, by contrast, fails to be diffeomorphism
invariant and the operator diverges on $\H^S_{\rm grav}$ in the
limit. Therefore, to obtain a well-defined operator, we used loops
enclosing an area $a_o$, the smallest non-zero quantum of area in
quantum geometry. The resulting operator can be regarded as a
`good' quantization of the classical constraint function because
it has the correct classical limit. The resulting quantum
constraint equation has novel and physically appealing properties.
First, it is a \emph{difference} --rather than differential---
equation and thus provides a `discrete time evolution'. Second,
the coefficients in this difference equation are such that the
`evolution' does \emph{not} break down at the singularity; quantum
physics does not stop at the big-bang! This occurs without fine
tuning matter or making it violate energy conditions. Furthermore,
while in consistent discrete models the singularity is often
`avoided' because discrete `time steps' are such that one simply
leaps over the point where the singularity is expected to occur
\cite{gp}, here, one can and does confront the singularity
\emph{head on} only to find that it has been resolved by the
quantum `evolution'. Furthermore, these features are robust
\cite{mb5,mb6}. However, near the big-bang, the state is
`extremely quantum mechanical,' with large fluctuations. Thus, the
classical space-time `dissolves' near the big-bang. In this regime, we
can analyze the structure only in quantum mechanical terms; we can
no longer use our classical intuition which is deeply rooted in
space-times and small fluctuations around them.

In the Planck regime, the predictions of loop quantum cosmology
are thus markedly different from those of standard quantum
cosmology based on geometrodynamics. The origin of this difference
can be traced back to the fact that while loop quantum cosmology
makes a crucial use of the fundamental discreteness of quantum
geometry, standard cosmology is based on a continuum picture. One
would therefore expect that the difference between the two would
become negligible in regimes in which the continuum picture is a
good approximation. We established two results to show that this
expectation is indeed correct. First, there is a precise sense in
which the difference equation of loop quantum cosmology reduces to
the Wheeler-DeWitt differential equation in the continuum limit.
Second, in the regime far removed from the Planck scale, solutions
to the Wheeler-DeWitt equation solve the difference equation to an
excellent accuracy. Thus, the quantum constraint of loop quantum
cosmology  modifies the Wheeler-DeWitt equation in a subtle
manner: the modification is significant only in the Planck regime
and yet manages to be `just right' to provide a natural resolution
of the big-bang singularity.

Next, let us re-examine the early papers on loop quantum cosmology
in terms of the precise mathematical framework developed in this
paper. In the present terminology, in the previous discussion one
effectively restricted oneself just to \emph{periodic} functions
$\exp (inc/2)$, rather than \emph{almost} periodic functions
$\exp (i\L c/2)$ considered here. Thus the gravitational Hilbert
space $\H^{S,P}_{\rm grav}$ considered there is the rather small,
periodic sub-space of the present $\H^S_{\rm grav}$. While this
restriction did have heuristic motivation, it amounted to forcing
$c$ to be periodic.%
\footnote{In the symmetry reduction, one began with the general
geometric theory of invariant connections and found that
components of a homogeneous connection transform as scalars under
gauge transformations \cite{mb9}. The appropriate, polymer theory
for real-valued scalar fields was developed only recently
\cite{als} and requires, as in the present paper, the Bohr
compactification of the real line.}
From the geometrodynamical perspective, the extrinsic curvature
was made periodic (with a very large period) and it was then not
surprising that the eigenvalues of the scale factor (and hence
also the volume) operator could only be discrete. However, a
careful analysis shows that the restriction to periodic functions
can not be justified: periodic functions fail to separate the
symmetric connections. Thus, in the earlier treatments, the space
of configuration variables was `too small' already at the
classical level and this led to an artificial reduction in the
size of the quantum state space.

In the analysis presented here, $c$ is \emph{not} periodic. As a
consequence, the spectrum of the volume operator is the entire
real line. Yet, there is discreteness in a more subtle sense: all
eigenvectors of the volume operator are normalizable. This is a
direct consequence of the fundamental premise of loop quantum
gravity that the quantum Hilbert space carries well defined
operators corresponding to \emph{holonomies} and not connections
themselves. In the full theory, this feature does make the spectra
of geometric operators discrete and their eigenvectors
normalizable. Because of the homogeneity assumption, however, the
first of these features is lost in loop quantum cosmology but the
second does survive. The surprising and highly non-trivial fact is
that this is sufficient for several of the main results of earlier
papers to continue to hold: i) the inverse scale factor is still
bounded from above; ii) the Hamiltonian constraint is again a
difference equation; and, iii) the coefficients in this equation
are such that the singularity is resolved in the quantum theory.
Furthermore, the current analysis provided a systematic approach
to verify that the constraint operator $\hat{C}^{(\L_o)}_{\rm
grav}$ has the correct classical limit and made its relation to
the Wheeler-DeWitt operator more precise and transparent. However,
as in earlier papers, the issue of finding the inner product on
physical states is yet to be analyzed in detail. While the group
averaging procedure \cite{dm,almmt,tt1} provides a natural avenue, a
detailed implementation of this program has only begun. The issue
of whether the `pre-classicality' condition selects unique quantum
states, thereby providing a natural solution to the issue of
initial conditions can be addressed systematically only after one
has a better control on the physical Hilbert space. Finally, the
discussion of section \ref{s4.3} not only shows that the classical
Einstein's equation is recovered in loop quantum cosmology in an
appropriate limit but it also provides a systematic approach to
the problem of finding quantum corrections to Einstein's
equations.%
\footnote{There appears to be a rather general impression that
Einstein's equations are not modified in loop quantum gravity. As our
discussion of \ref{s4.3} shows, this is \emph{not} the case.  It is
true that we simply promoted the classical Hamiltonian constraint
function to an operator. However, because there is no direct operator
analog of $c$, this `quantization' is subtle and even on
semi-classical (coherent) states, sharply peaked at classical
configurations, the expectation value of the constraint operator
equals the classical constraint function with small but very specific
quantum corrections. We expect that the same procedure can be applied
in the full theory to obtain quantum corrections to full Einstein's
equations.}
These corrections are now being worked out systematically.

We conclude with a general observation. The way in which the big-bang
singularity is resolved has potentially deep implications on questions
about the origin of the universe. For instance, the question of
whether the universe had a beginning at a finite time is now
`transcended'. At first, the answer seems to be `no' in the sense that
the quantum evolution does not stop at the big bang.  However, since
space-time geometry `dissolves' near the big-bang, there is no longer
a notion of time, or of `before' or `after' in the familiar sense.
Therefore, strictly, the question is no longer meaningful. The
paradigm has changed and meaningful questions must now be phrased
differently, without using notions tied to classical space-times.  A
similar shift of paradigm occurred already with the advent of general
relativity. Before Einstein, philosophers argued that the universe
could not have a finite beginning because, if it did, one could ask
what there was before. However, this question pre-supposes that
space-time is an eternal, passive arena and matter simply evolves in
it. With general relativity, we learned that space and time are `born
with matter', whence the question of `what was there before' is no
longer meaningful. Loop quantum cosmology brings about a further shift
of paradigm, weeding out certain questions that seemed meaningful in
classical general relativity and requiring that they be replaced by
more refined questions, formulated in the context of quantum
space-times.

\bigskip\bigskip

\textbf{Acknowledgments:}\,\, We thank the participants in the
`Quantum field theory on curved space-times' workshop at the Erwin
Schr\"odinger Institute for stimulating discussions. We are
especially grateful to Bill Unruh for raising several conceptual
issues addressed here, and to Klaus Fredenhagen for the key
observation that the Bohr compactification of the real line is the
appropriate quantum configuration space. This work was supported
in part by the NSF grant PHY-0090091, KBN grant 2P03B 2724, the
Eberly research funds of Penn State and the Erwin Schr\"odinger
Institute.

\appendix
\section{Method of Steepest Descent} \label{sa}

In this appendix, we will show that the method of steepest descent
can be used to evaluate the right side of the Poisson resummation
formula (\ref{poisson}):
\[
 \sum_n\, e^{-\epsilon^2(n-N)^2}f(n)\, =\, \sum_n\, \int
 e^{-\epsilon^2(y-N)^2}f(y)\, e^{2\pi iyn}\, dy\,.
\]
To begin with, we will assume that $f$ is analytic and return to
the cases of interest at the end.

Note first that the Fourier integral can be written as
\[
 \int e^{-\epsilon^2(y-N)^2}f(y)\, e^{2\pi iyn}dy= N\int
 g(z)\,e^{Nh_n(z)}\,dz \, ,
\]
where $g(z)=f(Nz)$ is the combination of volume eigenvalues, and
\[
 h_n(z):=-\epsilon^2N(z-1)^2+2\pi inz\,.
\]
For large $N$, the integral can be evaluated by the method of
steepest descent. For this, we have to first find the saddle
points. In our case, this amounts to finding solutions $z_0$ of
$h'(z_0)=-2\epsilon^2N(z_0-1) +2\pi in=0$. This equation has a
single solution, $z_0=1+i\pi n/N\epsilon^2$.
Steepest paths through $z_0$ are defined by constant imaginary part
${\rm Im}h_n(z)={\rm Im}h_n(z_0)$. If we write $z=\xi+i\eta$ with
real $\xi,\eta$, we obtain $-2\epsilon^2N(\xi-1)\eta+2\pi
n\xi=2\pi n$, which has the solutions $\xi=1$ or $\eta=\pi
n/\epsilon^2N$. We can now deform the original integration along
the real line to an integration along the steepest path $\eta=\pi
n/\epsilon^2N$, i.e.\ $z=t+i\pi n/\epsilon^2N$ with $t$ real.
Next, let us change the integration variable to $u$ by
$t=1+u/\epsilon\sqrt{N}$ such that
\[
 u^2=h_n(z_0)-h_n(z)=\epsilon^2N(t-1)^2\,.
\]
With $dz=du/\epsilon\sqrt{N}$ we obtain
\[
 \int g(z)\,e^{Nh_n(z)}\,dz =
 \epsilon^{-1}N^{-\frac{1}{2}} \int_{-\infty}^{\infty} g(t(u)+i\pi
 n/\epsilon^2N)\, e^{N(h_n(z_0)-u^2)}\, du\,.
\]

So far, everything was exact. In order to be able to compute the
integral, we now use the Taylor expansion:
\begin{eqnarray*}
 g(t(u)+i\pi n/\epsilon^2N) &=& g(1+u/\epsilon\sqrt{N}+i\pi
 n/\epsilon^2N)
 = g(1+i\pi n/\epsilon^2N)\\
 && +u g'(1+i\pi
 n/\epsilon^2N)/\epsilon\sqrt{N}+u^2 g''(1+i\pi
 n/\epsilon^2N)/\epsilon^2N \\
 &&+ O(u^3/\epsilon^3N^{\frac{3}{2}})
\end{eqnarray*}
Then, the integral can be evaluated approximately as \cite{ae}:
\begin{eqnarray*}
 \int g(z)\, e^{Nh_n(z)}\,dz &=& \epsilon^{-1}N^{-\frac{1}{2}}
 \left[g(1+i\pi n/\epsilon^2N) \int_{-\infty}^{\infty}
 e^{N(h_n(z_0)-u^2)}du\right.\\
 &&+ \epsilon^{-2}N^{-1} g''(1+i\pi n/\epsilon^2N)
 \int_{-\infty}^{\infty} u^2 e^{N(h_n(z_0)-u^2)}du\\
 && +\left.O(\epsilon^{-4}N^{-2}\int_{-\infty}^{\infty}
 u^4 e^{N(h_n(z_0)-u^2)}du)\right] \\
 &=& \sqrt{\pi}\epsilon^{-1}N^{-1} g(1+i\pi n/\epsilon^2N)
 e^{-\pi^2n^2/\epsilon^2+ 2\pi i nN}\\
 &&\times (1+ O((N\epsilon)^{-2}))\, ,
\end{eqnarray*}
where the corrections of order $(N\epsilon)^{-2}$ are small because
$d\ll N\L_o$. (We assumed $g''(z_0)$ to be of the same order as $g(z_0)$,
which is the case for the functions we are interested in here.)

We can now return to the Poisson re-summation formula and compute
the sum:
\begin{eqnarray*}
 \sum_ne^{-\epsilon^2(n-N)^2}f(n) &=& \sum_n \int
 e^{-\epsilon^2(y-N)^2}f(y)\, e^{2\pi iyn}\, dy\\
 & =& N\sum_n\int
 f(Nz)\, e^{Nh_n(z)}\, dz\\
 &=& \sqrt{\pi}\epsilon^{-1}\, \sum_n \, f(N+i\pi n/\epsilon^2)\,
 e^{-\pi^2n^2/\epsilon^2+ 2\pi inN}\\
 &&\times (1+ O((N\epsilon)^{-2}))\\
 &=& \sqrt{\pi}\epsilon^{-1} f(N) (1+ O(e^{-\pi^2/\epsilon^2})+
 O((N\epsilon)^{-2}))
\end{eqnarray*}
where, in the last step, we used the fact that terms with
$n\not=0$ are suppressed by the exponential because $\epsilon\ll
1$.

Finally, let us address the fact that the functions $f(y)$ of interest
in the main text come from the eigenvalues of the volume operator
which are of the type $V_{n\mu_o} = (\gamma |n|\mu_o/6)^{3/2}$ and,
because of the absolute value involved, $f(y)$ fail to be analytic in
$y$ at $y=0$. One can circumvent this potential problem by replacing
them with functions $\tilde{f}(y)$ which are the analytic
continuations (to the upper half complex $y$-plane) of the restriction
of $f(y)$ to the positive real $y$-axis and then carry out the above
calculation for $\tilde{f}$ which, by construction, are
analytic. (Thus, for example, $|y|^{3/2}$ is replaced by
$(y^2)^{3/4}$.) The error, $\sum_n \exp(-\epsilon^2
(n-N)^2)\, [f(n)-\tilde{f}(n)]$, can be
shown to be of the order $O(e^{ -(N\epsilon)^2})$ which is
negligible compared to the corrections derived above. The error is
so small because the function integrated on the right side of the
Poisson resummation formula is strongly peaked at a very large
positive value $y= N$ of $y$ and the contribution from the
negative half of the real $y$ axis is extremely small.

\end{document}